\documentclass[twocolumn,aps,showpacs,showkeys]{revtex4}

\usepackage{epsfig}
\usepackage{graphicx}

\usepackage{amssymb}

\begin{document}

\bibliographystyle{apsrmp} 

\title{Wide-band and Air Dispersion Effecting
the ABCD Algorithm of Phase-Recovery in Long-baseline Interferometry}
\author{Richard J. Mathar}
\affiliation{Leiden Observatory, Leiden University, Postbus 9513, 2300 RA Leiden, The Netherlands}
\email{mathar@strw.leidenuniv.nl}
\homepage{http://www.strw.leidenuniv.nl/~mathar}

\begin{abstract}
Long-baseline interferometry detects fringes created by superposition of two
beams of light collected by two telescopes pointing into a common direction.
The external path difference generated by pointing away from the zenith
is commonly compensated by adding a variable optical path length (delay)
through air for one beam such that the optical path difference
between the beams remains close to zero near the detector.

The ABCD formula assigns a (wrapped) phase to the amplitudes A to D of an
interference pattern shifted by multiples of 90 degrees in phase.
We study the interplay between a wide band pass of the optics and the
dispersion of the air in the compensating delay, which leads to small
deviations between the ABCD phase and the reduced, monochromatic
group-delay representation of the wave packets.

In essence, this adds dispersion to the effects that have been discussed
for evacuated interferometers (telescopes in space) before
[J.\ Opt.\ Soc.\ Am.\ A 22 (2005) 2774]. 
\end{abstract}

\maketitle
\section{Overview}
Angular astrometry with long-baseline interferometers
is based on planar trigonometry and measures baselines and path delays.
The metric of path delays may be derived
with sub-wavelength resolution if the fringe amplitude is fitted to a local,
monochromatic oscillation. This paper is an extended comment on the meaning
of the wavelength and phase of this fit. Astronomical interferometers
are generally built to accept large spectral band widths to enhance sensitivity
at low light levels; the terminology of ``mean'' and ``effective''
wavelengths must be attached to well-founded statistics (integrals) over
the broad-band spectra, in particular if the narrow-band concept of
the group delay is used to describe the propagation of light through delay
lines filled with air.

Chapter \ref{sec.wpack}
is a tutorial on the group delay concept of waves traveling through
a dispersive medium.
Chapter \ref{sec.synth}
is a review of the role played by the fringe phase to reconstruct
external path differences of star light approaching two telescopes (the
long-baseline interferometer). The case of mapping the entire K band 
of the near infrared on a single detector channel is used as a
numerical example for fringe patterns generated by a wide band pass.
Chapters \ref{sec.gDelay} and \ref{sec.abcdphase} describe a single channel
and an ABCD type of measurement of raw data and focus on the deconvolution
within the data reduction process which separates spectral intensities
(visibilities) from phases.
Chapter \ref{sec.kjitter} shows examples of the variation of the pivot (mean)
wave number---reduction of the mean correlated flux to a single representative
spectral line---as a function of star spectral indices and atmospheric
transmission.

\section{Wave Packets}\label{sec.wpack}
\subsection{Polychromatic Wave Trains}

\begin{figure}[hbt]
\includegraphics[width=8.5cm]{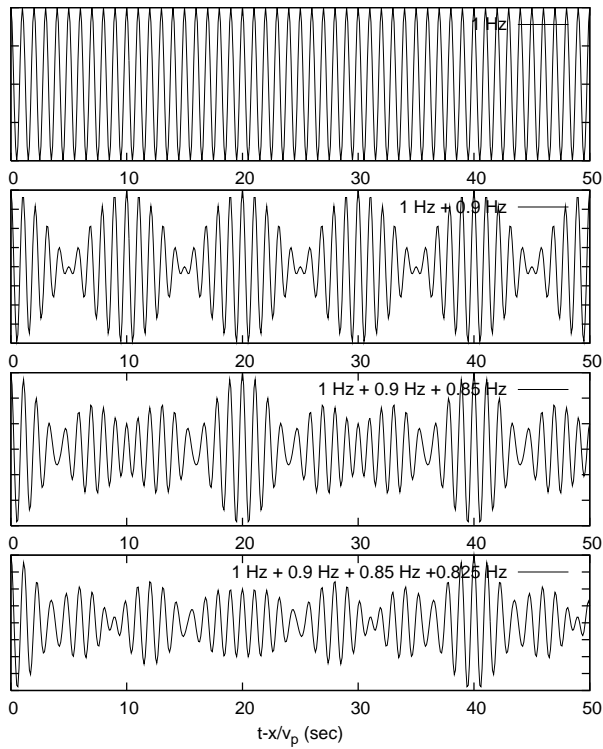}
\caption{
Top: a monochromatic wave at 1 Hz. Second from top: a superposition
of two waves of equal amplitude at 1 and 0.9 Hz. Below:
the sum of three waves at 1, 0.9 and 0.85 Hz. Bottom:
Four waves with differences of 0.1, 0.05 and 0.025 Hz in frequency.
}
\label{wpac.ps}
\end{figure}

The linear superposition of two electromagnetic fields of
angular frequencies
$\omega_1=v_p k_1$ and  $\omega_2=v_p k_2$
traveling at a common phase velocity $v_p$
generates frequencies at the mean and the difference,
\begin{eqnarray}
&&\cos(k_1 x-\omega_1 t)+\cos(k_2 x-\omega_2 t)
= \nonumber\\
&& \cos[k_1(x-v_p t)]+\cos[k_2(x-v_p t]
= \nonumber\\
&& 2\cos\left[\frac{\omega_1+\omega_2}{2}\left(\frac{x}{v_p}-t\right)\right]
  \cos\left[\frac{\omega_1-\omega_2}{2}\left(\frac{x}{v_p}-t\right)\right].
\end{eqnarray}
(This is rather arbitrarily written in terms of cosines, not sines, to simplify
embedding into a complex valued notation.)
The combined carrier frequency $(\omega_1+\omega_2)/2$ is close to the carrier
frequency of the individual components; the new aspect is the amplitude modulation
at the difference of the frequencies. Adding more waves of comparable magnitude
grows a few periodic pulses at the expense of increasingly more and more of the
low-frequent pulses. Figure \ref{wpac.ps} demonstrates
this bunching, starting with the example of a pulse train with a spacing of 0.1
Hz for the superposition of two frequencies at 1 and 0.9 Hz. The maxima with a
period of 20 s become dominant after adding 0.85 Hz, and the period of the
strongest maxima is eventually stretched to 40 s as the common divisor of the
frequencies becomes 0.025 Hz.

\subsection{Group Velocity}
The transition from a discrete spectrum to a continuous spectrum depletes
all pulses in the wave packet; essentially one remains with a carrier
frequency representing a mean of the frequencies in the wave packet,
and an envelope (hull) spreading inversely proportional to the
width of the spectrum. The apparent velocity $v_g$ of this wave packet
is defined by virtually marking some point with amplitude $S(x,t)$
at location $x$ and time $t$, waiting for a time interval $\Delta t$,
and measuring by which distance $\Delta x$ it has been moved:
\begin{equation}
S(x,t)=S(x+\Delta x,t+\Delta t).
\label{eq.Sgroup}
\end{equation}
The sum of the discrete polychromatic spectrum has become a Fourier integral
for the continuous case,
$S(x,t)=\int\frac{dk d\omega}{(2\pi)^2}e^{ikx-i\omega t}S(k,\omega)$.
If we use the condition of stationarity, Eq.\ (\ref{eq.Sgroup}), in the
Fourier representation,
\begin{eqnarray}
&&\int\frac{dk d\omega}{(2\pi)^2}e^{ikx-i\omega t}S(k,\omega)
= \nonumber \\
&&\int\frac{dk d\omega}{(2\pi)^2}e^{ikx-i\omega t}e^{ik\Delta x}e^{-i\omega\Delta t}S(k,\omega),
\label{eq.SFour}
\end{eqnarray}
we may expand $S(k,\omega)$ in the exponentials around pivot values
$k_0$ and $\omega_0$ representative of the center of the spectrum,
\begin{eqnarray}
e^{ik\Delta x-i\omega\Delta t} &\approx&
e^{ik_0\Delta x-i\omega_0\Delta t} \nonumber\\
&& \!\!\!\!\!\times \left[1+i(k-k_0)\Delta x-i(\omega-\omega_0)\Delta t+\cdots\right]
\label{eq.STayl}
\end{eqnarray}
The leading term in this expansion ensures that (\ref{eq.SFour})
is valid to the order that does not depend on $\Delta x$ and $\Delta t$. The
group velocity is defined as the ``observed'' motion
$v_g\equiv \lim_{\Delta t\rightarrow 0} \Delta x/\Delta t$.
To fulfill (\ref{eq.STayl}) also to the linear order $O(\Delta x,\Delta t)$,
the eliminating constraint becomes
\begin{equation}
i(k-k_0)\Delta x-i(\omega-\omega_0)\Delta t=0.
\label{eq.dxdt}
\end{equation}
Division through $i\Delta t$ and $i(k-k_0)$
means \citep{MatharAJP66}\cite[\S 1.3.4]{Born}
\begin{equation}
v_g=\frac{\omega-\omega_0}{k-k_0}\equiv\frac{\partial\omega}{\partial k}. \label{eq.vg}
\end{equation}
The transition from the quotient to the differential implies
that the band width is small in the sense that (i) the additional
variation
of $S(k,\omega)$ in the kernel of (\ref{eq.SFour}) is negligible,
and (ii) the second and higher order derivatives of the eigenmodes
$\omega(k)$ are also negligible.

\subsection{Group Refractive Index}\label{sec.vgdef}
The microscopic absorption and emission processes of an electromagnetic
wave passing through gaseous or condensed matter can be summarized as a
wavelength dependent refractive index $n(k,\omega)$; in the long-wavelength
limit---i.e., $\lambda\equiv 2\pi/k$ much larger than the inter-molecular
distance or lattice unit cell---the superposition of many of these elementary
processes replaces the vacuum wavelength by eigenmodes of the dispersion
\begin{equation}
ck=n(k,\omega)\omega . \label{eq.ndef}
\end{equation}
Division through $kn$ replaces the vacuum phase velocity $c$ by the
phase velocity
\begin{equation}
v_p\equiv \frac{c}{n}=\frac{\omega}{k}. \label{eq.vphase}
\end{equation}

The concept of the group refractive index transforms this pair of
variables $v_p$ and $n$ further to a pair of variables $v_g$ and $n_g$ to
acknowledge that the relevant speed of fringe packets switches from $v_p$
to $v_g$: If the setup is concerned with moving them through
different geometric paths and monitoring a high coherent signal at a detector
position, the geometric path lengths must involve $v_g$ to synchronize
the times of arrival to achieve interference near the detector.
To switch from $(v_p,n)$ to $(v_g,n_g)$, (\ref{eq.vphase})
is rewritten to define the group refractive index $n_g$,
\begin{equation}
v_g\equiv \frac{c}{n_g}. \label{eq.ngdef}
\end{equation}
The derivative of (\ref{eq.ndef}) w.r.t.\ $k$ is with (\ref{eq.vg})
\begin{equation}
c=\frac{\partial n}{\partial k}\omega+n \frac{\partial d\omega}{\partial k}=
\frac{\partial n}{\partial k}\frac{ck}{n}+n v_g.
\end{equation}
This equation is solved for $v_g$ and inserted into (\ref{eq.ngdef}),
\begin{equation}
n_g=\frac{c}{v_g}=\frac{c}{\frac{c}{n}-\frac{\partial n}{\partial k}\frac{ck}{n^2}}
=\frac{n}{1-\frac{\partial n}{\partial k}\frac{k}{n}}.
\end{equation}
This equation is accurate, see Eqs.\ (4)--(5) in \citep{DaigneAApS138}
and Eq.\ (3) in \citep{HarunaOL23}. Assuming weak dispersion,
its geometric series expansion in powers of the small $\partial n/\partial k$
may be truncated after the linear term:
\begin{equation}
n_g\approx n\left(1+\frac{\partial n}{\partial k}\frac{k}{n}\right)
=n+k\frac{\partial n}{\partial k}, \label{eq.vgapprox}
\end{equation}
which is the version in \cite[p.\ 1666]{CiddorAO38} and \cite[p.\ 768]{PavlicekAO43}.
The relative error in (\ref{eq.vgapprox}) is approximately the first
neglected order,
$(\frac{\partial n}{\partial k}\frac{k}{n})^2
=
(\frac{\partial n}{\partial \sigma}\frac{\sigma}{n})^2
$
where $\sigma=k/(2\pi)=1/\lambda$ is the spectroscopic wave-number.
In the K band around $\sigma\approx 4400$ cm$^{-1}$
at air pressures around 750 hPa,
a typical value is $\partial n/\partial \sigma\approx 1.2\times 10^{-10}$ cm,
which yields a relative error $\approx 2\times 10^{-13}$.
In the N band \citep{MatharAO43,ColavitaPASP116,HillIP26,HillJOSA70},
the relative error would still be tiny.

A short-hand with one (and more) primes as in
\begin{equation}
n'(k) = \partial n/\partial k
\end{equation}
will subsequently be used to represent the first (and higher) order derivatives
w.r.t.\ $k$.

\subsection{Time Delay}
A group delay $t_g$ is implicitly defined
via (\ref{eq.dxdt}) as the time $\Delta t$ for a \emph{fixed} $\Delta x$.
Again in the limit of small deviations from a reference $(k_0,\omega_0)$, this
becomes 
\begin{equation}
t_g\equiv \Delta x \frac{\partial k}{\partial\omega}.
\end{equation}
The derivative of (\ref{eq.vphase}), $k=\omega/v_p$, with respect to
$\omega$ is $\partial k/\partial\omega=1/v_p-(\omega/v_p^2)(\partial v_p/\partial\omega)$
and inserted, using $t=\Delta x/v_p$ for the un-dispersed reference time delay,
\begin{equation}
t_g=\Delta x \frac{1}{v_p}-\Delta x\frac{\omega}{v_p^2}\frac{\partial v_p}{\partial\omega}
= t-\omega \frac{t}{v_p}\frac{\partial v_p}{\partial \omega}.
\label{eq.tginterm}
\end{equation}
At fixed $\Delta x$, the total differential of $x=tv_p$ must remain zero,
$t dv_p+v_pdt=0$, which is used to substitute $t\partial v_p\leadsto -v_p\partial t$
in (\ref{eq.tginterm}) and to recover the
Meisner
formula \citep{MeisnerSPIE5491}
\begin{equation}
t_g=t+\omega\frac{\partial t}{\partial \omega}.
\end{equation}

\section{Synthetic Broad-band Fringes}\label{sec.synth}
\subsection{Model Air Dispersion}
\begin{figure}[hbt]
\includegraphics[width=8.5cm]{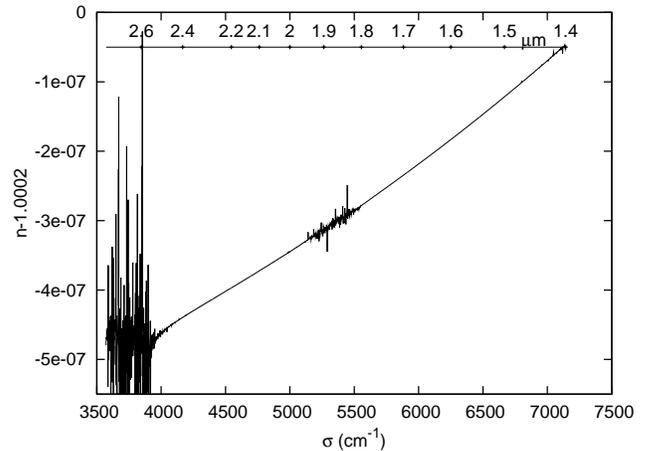}
\caption{
A prototypical infrared dispersion of humid air
at $p=744$ hPa, $T=16$ $^\circ$C and 10\% humidity
in the K band around 4500 and the H band around 6500 cm$^{-1}$.
}
\label{nHK.ps}
\end{figure}

In the numerical examples that follow, a model of the refractive index $n$
in the near-infrared is needed. It is demonstrated in Fig.\ \ref{nHK.ps}
calculated from an update of an oscillator strength summation \citep{MatharAO43}.
Since
we also employ integrations over the dispersion spectrum, it is favorable
to replace this detailed spectrum by a fit,
explicitly
\begin{eqnarray}
n-1 &=& \sum_{i=0,1,\ldots} c_i(T,p,H) \left(\sigma - \sigma_\mathrm{ref}\right)^i ;
\\
c_i(T,p,H) &=& c_{i\mathrm{ref}}
+ c_{iT}\left( \frac{1}{T}-\frac{1}{T_\mathrm{ref}}\right)
\nonumber \\
&& + c_{iH}\left(H-H_\mathrm{ref} \right)
+ c_{ip}\left(p-p_\mathrm{ref} \right).
\label{eq.fit}
\end{eqnarray}
Here, $T$ is the absolute temperature with a reference value of $T_\mathrm{ref}$,
$p$ is the air pressure with a reference value set
at $p_\mathrm{ref}$, $H$ the relative humidity between 0 and 100 with a reference
value set at $H_\mathrm{ref}$, $\sigma$ is the wave-number with
a reference value set at $\sigma_\mathrm{ref}$,
and $c_i$ are the expansion coefficients of Table \ref{tab.fitn_1}.

\begin{figure}[hbt]
\includegraphics[width=8.5cm]{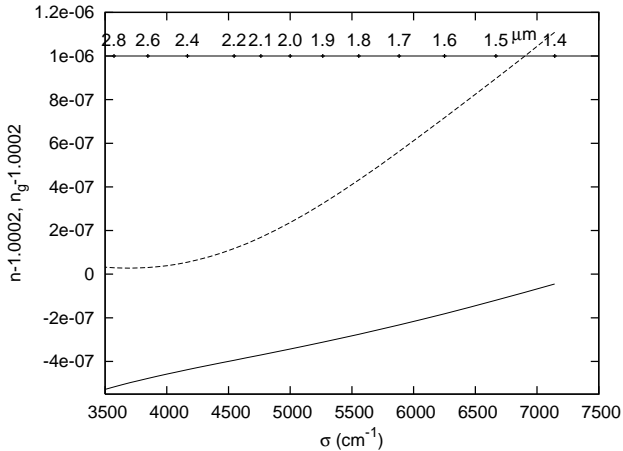}
\caption{The lower line is $n$ fitted to Fig.\ \ref{nHK.ps}
according to (\ref{eq.fit}); the upper line is $n_g=n+n'k$
derived from this fit according to (\ref{eq.vgapprox}).
}
\label{n.ps}
\end{figure}

\begin{table}[hbt]
\begin{tabular}{c|cc}
$i$ & $c_{i\mathrm{ref}}$ / cm$^i$ & $c_{iT}$ / [cm$^i$K]
   \\
\hline
0 & $0.000199594$ 
  & $0.0583358$
\\
1 & $0.113225\times 10^{-9}$ 
  & $-0.330288\times 10^{-7}$ 
\\
2 & $-0.438053\times 10^{-14}$ 
  & $0.811015\times 10^{-10}$ 
\\
3 & $0.101921\times 10^{-16}$
  & $-0.514801\times 10^{-13}$
\\
4 & $-0.296872\times 10^{-20}$
  & $0.150602\times 10^{-16}$
\\
5 & $0.311755\times 10^{-24}$
  & $-0.157482\times 10^{-20}$
\\
\end{tabular}

\begin{tabular}{c|cc}
$i$ 
&  $c_{iH}$ / [cm$^i$/\%]
& $c_{ip}$ / [cm$^i$/Pa]
\\
\hline
0 
 & $ -0.904550\times 10^{-8}$
& $0.268437\times 10^{-8}$ 
\\
1 
& $ 0.119196\times 10^{-11}$
& $0.136651\times 10^{-14}$ 
\\
2 
& $-0.149163\times 10^{-14}$
& $0.136002\times 10^{-18}$ 
\\
3
  & $0.985685\times 10^{-18}$
& $0.822304\times 10^{-23}$
\\
4
  & $-0.288288\times 10^{-21}$
& $-0.224090\times 10^{-26}$
\\
5
  & $0.301528\times 10^{-25}$
& $0.251260\times 10^{-30}$
\\
\end{tabular}
\caption[]{
Fitting coefficients for the multivariate Taylor expansion (\ref{eq.fit})
to the real part of the index of refraction for
$T_\mathrm{ref}=(273.15+16)$ K,
$p_\mathrm{ref}=74400$ Pa,
$H_\mathrm{ref}=10$ \%,
$\sigma_\mathrm{ref}=1/(2.25\mu\mathrm{m})=4444.\bar 4$ cm$^{-1}$,
1.3 $\mu$m $<=1/\sigma<=$ 2.5 $\mu$m.
}
\label{tab.fitn_1}
\end{table}
The coefficient $c_{0T}$ means a change of the temperature by $\Delta T=1$ K
changes $\Delta \frac{1}{T}=-\frac{\Delta T}{T^2}$ by 10$^{-5}$ K$^{-1}$,
changes $n$ by $6\times 10^{-7}$ and relocates the wave packet by 60 $\mu$m
after a path of 100 m
if $\lambda=2.25$ $\mu$m. The coefficient $c_{0H}$ means a change
of the relative humidity by 10 percent points changes $n$ by $0.9\times 10^{-7}$
and relocates the packet by 9 $\mu$m over the same path at
the same wavelength.
Introducing the fit replaces Fig.\ \ref{nHK.ps} by Fig.\ \ref{n.ps}.

\subsection{Model of Two-beam Interference}\label{sec.emkmodl}
Illuminated by a source with coherent spectral energy density $e(k)$,
the superposition of the light with a second beam forced to a detour
of a compensating geometric path length $D_i$ yields a fringe pattern
\begin{equation}
A(D_i)=\int e(k)\{1 + \cos[ n(k)kD_i-kD_e ]\} dk,
\label{eq.Idfring}
\end{equation}
assuming no correlation between the fields (amplitudes) at different $k$.
$D_e$ is the (vacuum) delay imprinted on the two beams at arrival of
the two telescopes, essentially the dot product between the baseline
vector and a unit vector into the pointing direction \citep{Boden}.
(The wavelength dependence of $D_e$ from lensing effects by the
atmosphere above the telescopes \citep{MatharBA14,MatharArxiv04,BoehmVLBI03_131}
and piston fluctuations are ignored here.)

\begin{figure}[hbt]
\includegraphics[width=8.5cm]{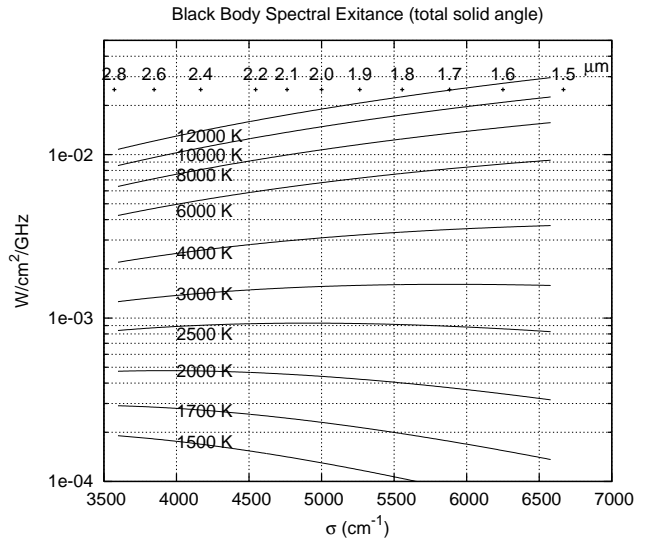}
\caption{
Black body spectral densities
$2\pi h\nu^3/[c^2(e^{h\nu/k_BT}-1)]$
between 1.52 and 2.78 $\mu$m
on a semi-logarithmic scale, where $h$ are the Planck
and $k_B$ the Boltzmann constant.
}
\label{blackb_tab.ps}
\end{figure}

\begin{figure}[hbt]
\includegraphics[width=8.5cm]{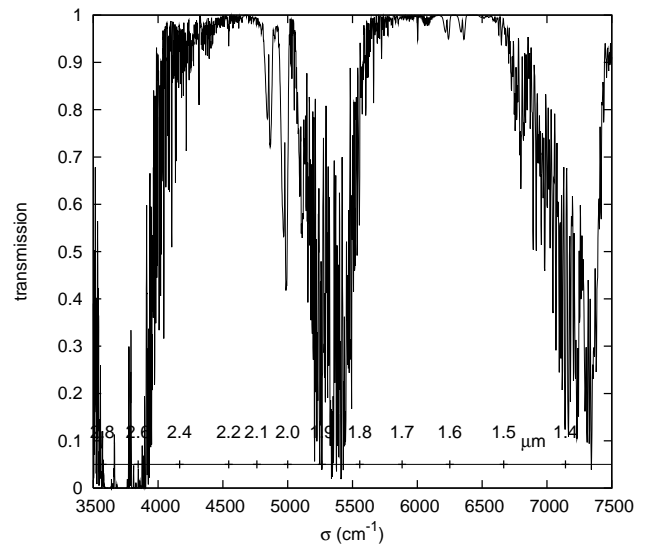}
\caption{
Synthetic Mauna Kea sky transmission in the K band at an airmass of 1.0
and a 1.6 mm water vapor column  \citep{LordNASA1992}.
}
\label{trans.ps}
\end{figure}

\begin{figure}[hbt]
\includegraphics[width=8.5cm]{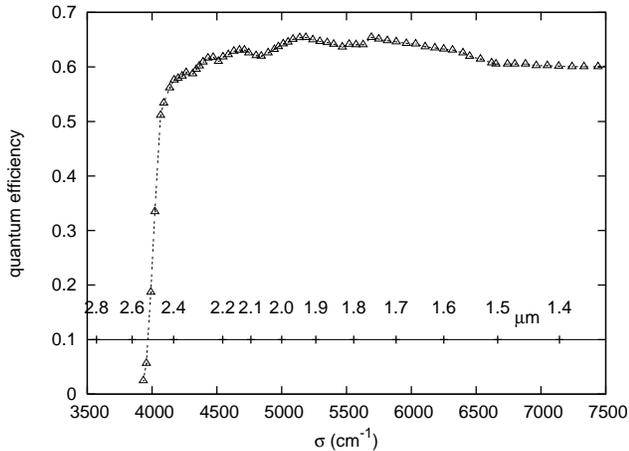}
\caption{
Quantum efficiency of the 256$\times$256 Rockwell PICNIC detector.
}
\label{256QE.ps}
\end{figure}

$D_i$ is the ``internal'' geometric path difference added by the mirror
train on the ground, and $nD_i$ the associated optical path difference,
which is controlled as to steer the phase of the cosine to some
desired value in some average sense.
The coherent flux $e(k)$ is the one experienced by the detector which produces
the signal $A$ \citep{StubbsApJ646}: it is the product of (i) the star emission,
by (ii) the Earth's sky transmission \citep{GandhiMNRAS337},
by (iii) the reflectivity
of mirror optics and transmissivity of beam combiner optics,
including vibrations and diffraction effects \citep{PuJopt24},
and by (iv)
the coupling efficiency of any fiber optics and quantum efficiency of the
detector, including tilt effects \citep{AitAmeurOptComm233,AslundOptComm262}.
We will represent the star emission  by black body spectra of Fig.\ \ref{blackb_tab.ps},
the sky transmission by the model of Fig.\ \ref{trans.ps}, ignore  chromaticism
of the transmission of intermediate optics and use the quantum efficiency
of Fig.\ \ref{256QE.ps}.

\begin{figure}[hbt]
\includegraphics[width=8.5cm]{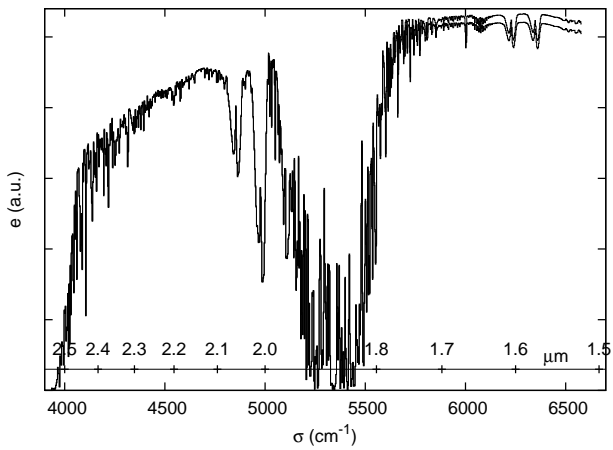}
\caption{
Prototypical model of $e(k)$ constructed as a product of Fig.\ \ref{trans.ps},
Fig.\ \ref{256QE.ps} and a black body of 8000 K or 9000 K.
}
\label{spektr.ps}
\end{figure}

A different approach to delay compensation establishes the term $nD_i$
with a dispersion compensator \citep{BergerSPIE4838,LevequeASS239,BrummelaarAO34}.
This trades the variability
of the air dispersion \citep{HillJOSA70} by the dependence of the glass
dispersion on temperature and expansion coefficients \citep{Brachet05}
and is not discussed here.

As a model of the band pass [specification of the integration interval
in (\ref{eq.Idfring})] we will use parameters that follow from
placing an Infrasil \citep{HeraeusInfrasil} prism of a wedge angle of 11.9$^\circ$ in
front of the detector, and reverse interpolation of the formula of the
refraction angle \citep{GreisenAAp446,Born}\cite[\S 4.7.2]{Born}.
An extreme setup in terms of band pass width would map the entire K band from
2.05 to 2.45 $\mu$m onto a single detector pixel with output $A$.
The non-linearity of the prism equation and prism material dispersion
would map most of the H band onto one of the side channels, as summarized
in Table \ref{tab.morePix}.

\begin{table}[hbt]
\begin{tabular}{cc|c|ccccc}
wedge ($^\circ$) & incid. ($^\circ$) & \# pixels & \multicolumn{5}{c}{limits ($\mu$m)} \\
\hline
11.9 & 8.4522 & 3 & 1.52 & 2.05 & 2.45 & 2.78 \\
22.59 & 10.0084 & 2 & 2.05 & 2.26 & 2.45 \\
31.7 & 14.3783 & 3 & 2.05 & 2.19 & 2.32 & 2.45 \\
39.04 & 18.173 & 4 & 2.05 & 2.15 & 2.25 & 2.35 & 2.45 \\
\hline
\end{tabular}
\caption{The spectral range
of detector pixels if the wedge angle of the prism
is increased for higher dispersion, and the incidence angle is chosen to
stay with a minimum total deflection angle between the direction of incidence
and the direction of exit. The calculations are for Infrasil assuming that
each pixel is equivalent to 0.0014077 rad difference in the exit angles.
\label{tab.morePix}
}
\end{table}

The main advantage of wide-band detection is the increase of the signal $A$
given limited intensity $e$ and noise by photon noise, detector readout and
thermal background. The focus of this work here is to point at some
inherent disadvantages.

\subsection{Delay Tracking}
The philosophy of coherencing is to actuate the delay line (DL),
the variable $D_i$, to manipulate the phase difference
\begin{equation}
\varphi(k)=nkD_i-kD_e
\label{eq.phi}
\end{equation}
in (\ref{eq.Idfring}).
Tracking on a local maximum (peak) of $A$ would try to achieve a broad maximum
or minimum of $\varphi$ as a function of $k$, since the nearby spectral elements
of $e(k)$ then all add up ``in phase.''
The difference between nulling the phase and this coherencing
is symbolized by
the difference between Fig.\ \ref{phase.ps} and Fig. \ref{phaseH.ps}:
\begin{figure}[hbt]
\includegraphics[width=8.5cm]{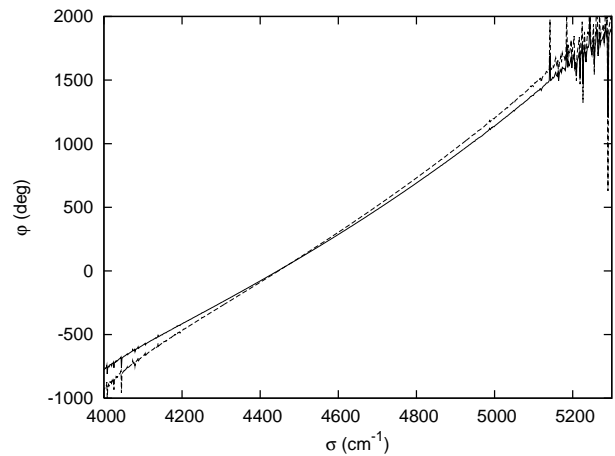}
\caption{
The residual phase $\varphi(k)$ for a geometric delay $D_e=100$ m
as a function of $\sigma=k/(2\pi)$:
The solid line is for 10\% humidity at 744 hPa, 16$^\circ$C and
$D_i=99.980\, 044\, 790$ m.
The dashed line for 20\% humidity at the same pressure and temperature
and $D_i=99.980\, 053\, 689$ m. The $D_i$ are chosen as to achieve
a common zero at $\lambda=2.25$ $\mu$m ($\sigma=4444$ cm$^{-1}$);
the 9 $\mu$m difference between the $D_i$ corresponds to $c_{0H}$ of
Table \ref{tab.fitn_1}.
}
\label{phase.ps}
\end{figure}
\begin{figure}[hbt]
\includegraphics[width=8.5cm]{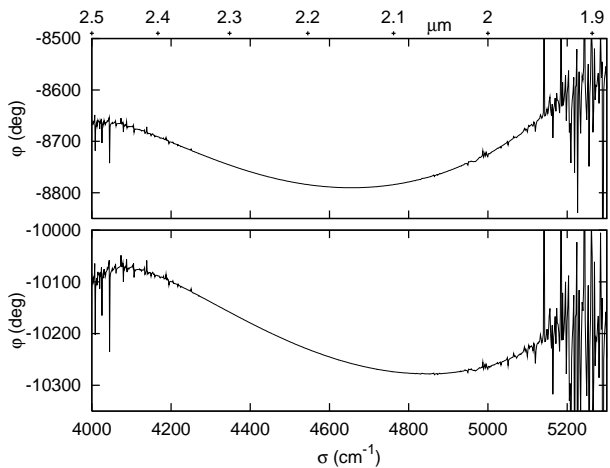}
\caption{
The phase $\varphi$ of (\ref{eq.phi}) for a geometric delay $D_e=100$ m
as a function of wave-number:
The upper plot is for 10\% humidity at 744 hPa, 16 $^\circ$C and
$D_i=99.97999$ m.
The lower plot for 20\% humidity at the same pressure, temperature
and $D_i$.
}
\label{phaseH.ps}
\end{figure}
Setting $D_i$ to steer $\varphi(k)$ close to zero as in Fig.\ \ref{phase.ps}
accumulates phases in the integral (\ref{eq.Idfring})
that differ by a few hundred degrees if $e(k)$ covers a broad range of the
spectrum. To achieve less chaotic/oscillatory behavior of the integrand, one
would move $D_i$ to achieve a flatter $\varphi(k)$ to reach
Fig.\ \ref{phaseH.ps} (see \citep{TubbsSPIE5491} for N band examples).
The spectral elements in a small interval of similar $\varphi(k)$ interfere
positively. The extremum of $\varphi(k)$ pinpoints the ``effective'' wave-number
because the populated density of $\varphi$-states is largest there.
Mathematically speaking, the extremum is characterized by a vanishing
derivative of (\ref{eq.phi}),
\begin{equation}
\frac{\partial \varphi}{\partial k}=n'kD_i+nD_i-D_e=0.
\label{eq.phiDeriv}
\end{equation}
Solving this 
for the geometric compensating path $D_i$ with (\ref{eq.vgapprox}) yields
\begin{equation}
D_i=D_e/n_g .
\label{eq.Digroup}
\end{equation}
This verifies the claim of Section \ref{sec.vgdef} that $n_g$ is the scale
factor to replace $n$ in the realm of fringe tracking.
In the limit of constant $n_g(k)$ and $n(k)$, the fringe pattern $A$ of
(\ref{eq.Idfring}) is an ordinary Fourier Cosine Transform of the
source spectrum $e(k)$, its Michelson interferogram.

Apart from a sign flip and scaling with $D_e$, Fig.\ \ref{phaseH.ps}
is equivalent to the upper part of \cite[Fig 1]{DaigneAApS138}:
Their fringe tracking selects a minimum of Fig.\ \ref{phaseH.ps}
by moving $D_i$, then chooses another (but unique) $\varphi_\mathrm{gui}$ in the
two adjacent spectral channels to define ``guiding'' wave-numbers.
However, Fig.\ \ref{phaseH.ps} points at a prospective problem
with large band passes: such a pair of wave-numbers may not exist in the
two neighboring channels when nonlinear dispersion, i.e.,
$n''(k)\neq 0$, may create a non-parabolic shape of $\varphi(k)$ such that
no two solutions to
$\varphi(k)=\varphi_\mathrm{gui}$ may exist due to the additional extremum.

\subsection{Numerical Examples}
\begin{figure}[hbt]
\includegraphics[width=8.5cm]{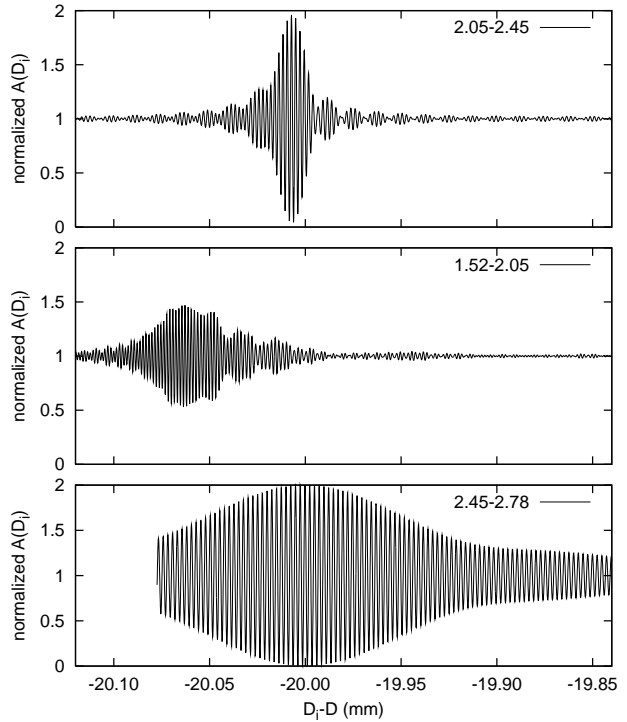}
\caption{
Normalized K band fringe $A(D_i)/e_0$ according to Eq.\ (\ref{eq.Idfring}) for the
8000 K spectrum of Fig.\ \ref{spektr.ps} for the three bands
of the first line in Table \ref{tab.morePix}
at an external delay of $D_e=100$ m,
including nonlinear air dispersion $n(k)$.
}
\label{fringe100.ps}
\end{figure}

\begin{figure}[hbt]
\includegraphics[width=8.5cm]{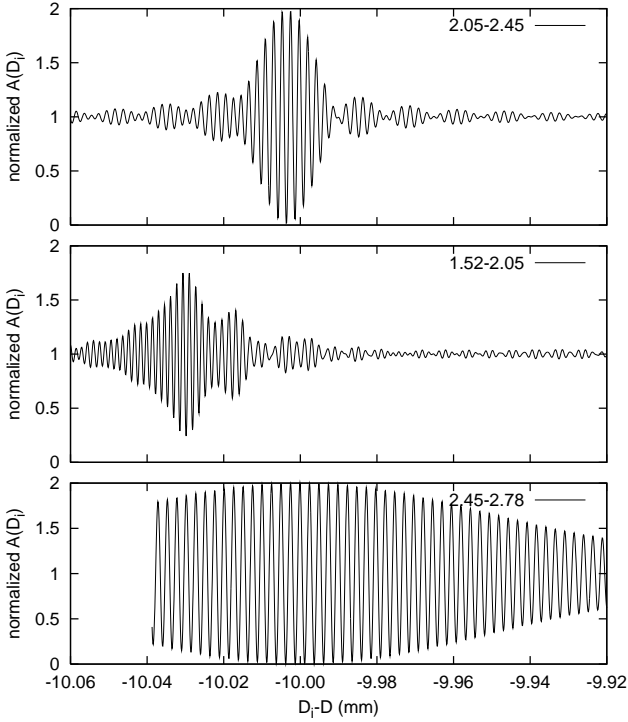}
\caption{
The calculations of Fig.\ \ref{fringe100.ps}
repeated for a vacuum delay of $D_e=50$ m.
}
\label{fringe50.ps}
\end{figure}

Fig.\ \ref{fringe100.ps} and \ref{fringe50.ps} show examples
of the interplay of the 8000 K spectrum of Fig.\ \ref{spektr.ps}
with the dispersion of Fig.\ \ref{n.ps}, integrated over three
adjacent intervals as specified by the first line of Table \ref{tab.morePix}.
The middle plot illustrates how the deep cut in the
spectrum from 1.8 to 1.9 $\mu$ initiates a break-up of the
main envelope into fragments in some sort of restoration
of the multiple packets scheme that was the foundation of Fig.\ \ref{wpac.ps}.

The roughest estimate of the position $D_i$ of the fringe center---ignoring
the effect of the weighting with the star color $e(k)$---is where the
argument of the cosine of Eq.\ (\ref{eq.Idfring}) becomes zero.
If one
assumes a constant refractive index $n(\bar k)$ inside the band, this is
\begin{equation}
D_i=D_e/n(\bar k).
\label{eq.Destim0}
\end{equation}
The relative error of this formula in comparison with the more accurate
(\ref{eq.Digroup}) is $n_g(\bar k)-n(\bar k)$, typically the distance
of $4\times 10^{-7}$ between the curves in Fig.\ \ref{n.ps}.

\section{Beyond the Group Delay Approximation}\label{sec.gDelay}
\subsection{Moment Expansion}\label{sec.emomen}
\subsubsection{Nonlinear Dispersion}
Up to here, we have focused on the dispersion $n(k)$ as the
key parameter to locate the fringe packet. In practise, the
``astrometric'' interpretation of the interferometric datum $A$
is concerned with the recovery of the external delay $D_e$ given $A$ and $D_i$
as the raw data. (In contrast to the ``imaging'' experiment which tries to measure
the magnitude of $A$ as a function of $D_e$, this quest is complementary
as it aims at determining \emph{where} on the $D_e$ axis the zero optical
path difference is, rather than determining the degree of coherence that 
would represent the star's brightness distribution.)
The key subject of this paper is the additional role of the
spectrum $e(k)$ and its interplay with the dispersion when
the weighting with the correlated flux lets the ``observed''
phase of $A(D_i)$ deviate from the ``guiding'' phase of an 
extremum of $\varphi(k)$.

To understand this phenomenon, one introduces a Taylor expansion
of the dispersion expansion around some ``central'' value $\bar k$
in the spirit of Filon quadratures \citep{IserlesPRS461},
\begin{eqnarray}
n(k)&=&n(\bar k)+(k-\bar k)n'(\bar k)+\frac{(k-\bar k)^2}{2} n''(\bar k) \nonumber \\
&& +\frac{(k-\bar k)^3}{6} n'''(\bar k)+\cdots.
\end{eqnarray}
Higher order derivatives are
\begin{eqnarray}
n_g(\bar k) & \equiv& n(\bar k)+\bar k n'(\bar k), \label{eq.ngofn}\\
n_g'(\bar k)\equiv \frac{\partial n_g}{\partial k}_{\vert\bar k}
&=& 2n'(\bar k)+\bar k n''(\bar k), \label{eq.ngprimeofn} \\
n_g''(\bar k) &=& 3n''(\bar k)+\bar k n'''(\bar k). \label{eq.ngdprime}
\end{eqnarray}
Quick calculations use Table \ref{tab.fitn_1}:
\begin{equation}
n_g'=\frac{1}{2\pi}\left(2\frac{dn}{d\sigma}+\sigma\frac{d^2n}{d\sigma^2}\right)
\end{equation}
evaluated at $\sigma_\mathrm{ref}$ is
\begin{eqnarray}
n_g'&=&\frac{1}{2\pi}\left(2c_\mathrm{1ref}+2\sigma_\mathrm{ref}c_\mathrm{2ref}
\right)\nonumber \\
&=&\frac{1}{\pi}(0.113225\times 10^{-9}-4444\times0.438\times 10^{-14})
\mathrm{cm} 
\nonumber \\
&\approx& 3.0\times 10^{-13} \mathrm{m},
\label{eq.ngprimeref}
\end{eqnarray}
for example.
Equivalent expansion of $e(k)$ defines the moments of the energy spectrum,
\begin{equation}
e_m\equiv \int e(k)(k-\bar k)^m dk
\end{equation}
around any chosen center $\bar k$ \citep{TuryshevAO42}.
For pure black body spectra, $e_m$ is a binomial sum over Debye functions up
to order $3+m$ \cite[\S 27.1]{AS}.

\subsubsection{Asymmetric Coherent Spectra}
Eq.\ (\ref{eq.Idfring}) is accordingly expanded up to third order
in $k-\bar k$,
\begin{eqnarray}
A(D_i)
\approx
\left\{ 1 + \cos[ n(\bar k)\bar kD_i-\bar kD_e ]\right\} e_0\nonumber \\
 - \left\{ [n'(\bar k) \bar k + n(\bar k)]D_i-D_e\right\}\sin[ n(\bar k)\bar kD_i-\bar kD_e ] e_1 \nonumber \\
 -\frac{1}{2} \left\{ [n'(\bar k) \bar k + n(\bar k)]D_i-D_e\right\}^2
\cos[ n(\bar k)\bar kD_i-\bar kD_e ]
e_2 \nonumber \\
 -\frac{1}{2} [n''(\bar k) \bar k + 2n'(\bar k)]D_i
\sin[ n(\bar k)\bar kD_i-\bar kD_e ]
e_2 \nonumber \\
 -\frac{1}{2} \left\{ [n'(\bar k) \bar k + n(\bar k)]D_i-D_e\right\}
 [n''(\bar k) \bar k + 2n'(\bar k)] \nonumber \\
\times D_i
\cos[ n(\bar k)\bar kD_i-\bar kD_e ]
e_3 \nonumber \\
 +\frac{1}{6} \left\{ [n'(\bar k) \bar k + n(\bar k)]D_i-D_e\right\}^3
\sin[ n(\bar k)\bar kD_i-\bar kD_e ]
e_3 \nonumber \\
 -\frac{1}{6} [n'''(\bar k) \bar k + 3 n''(\bar k)]D_i
\sin[ n(\bar k)\bar kD_i-\bar kD_e ]
e_3 +O(e_4)
\end{eqnarray}
\begin{eqnarray}
&\approx&
\left[ 1 + \cos \varphi(\bar k)\right] e_0
 - \left[ n_g(\bar k) D_i-D_e\right]e_1 \sin \varphi(\bar k)
\nonumber \\
&&
 -\frac{1}{2} \left[ n_g(\bar k) D_i-D_e\right]^2
e_2 \cos \varphi(\bar k)
 \nonumber \\
&&
 -\frac{1}{2} n_g'(\bar k) D_i
e_2 \sin \varphi(\bar k)
 \nonumber \\
&&
 -\frac{1}{2} \left[ n_g(\bar k) D_i-D_e\right]
 n_g'(\bar k) D_i e_3 \cos \varphi(\bar k)
 \nonumber \\
&&
 +\frac{1}{6} \left[ n_g(\bar k)D_i-D_e\right]^3
e_3 \sin \varphi(\bar k)
 \nonumber \\
&&
 -\frac{1}{6} n_g''(\bar k) D_i
e_3 \sin \varphi(\bar k)
 +O(e_4).
\label{eq.Idgen}
\end{eqnarray}
If we gather cosine and sine coefficients in the form
\begin{equation}
A=e_0+C\cos\varphi(\bar k)-S\sin\varphi(\bar k)
\end{equation}
with
\begin{eqnarray}
C &\equiv&
e_0
 -\frac{1}{2} \left[ n_g(\bar k) D_i-D_e\right]^2
e_2  \nonumber \\
&& -\frac{1}{2} \left[ n_g(\bar k) D_i-D_e\right]
 n_g'(\bar k) D_i e_3 +\cdots, \label{eq.Cdef} \\
S& \equiv &
  \left[ n_g(\bar k) D_i-D_e\right]e_1
+\frac{1}{2} n_g'(\bar k) D_i
e_2  \nonumber \\
&& -\frac{1}{6} \left[ n_g(\bar k)D_i-D_e\right]^3
e_3
 +\frac{1}{6} n_g''(\bar k) D_i
e_3+\cdots, \label{eq.Sdef}
\end{eqnarray}
the fringe packet is cast into a monochromatic format,
\begin{equation}
A = e_0+\sqrt{C^2+S^2}\cos[\varphi(\bar k)+\varphi_i],
\label{eq.Aampl}
\end{equation}
and rigorously defines an ``observed'' phase $\varphi(\bar k)+\varphi_i$,
\begin{equation}
\tan \varphi_i =S/C.
\label{eq.effphi}
\end{equation}

\subsubsection{Remark on Nomenclature}
\begin{figure}[hbt]
\includegraphics[width=8.5cm]{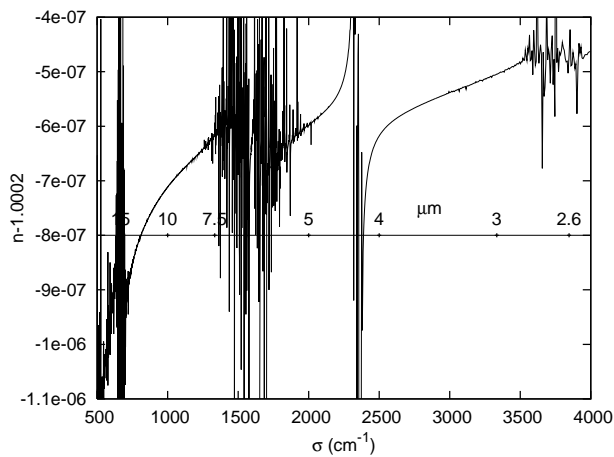}
\caption{The continuation of Fig.\ \ref{n.ps} into the
mid-infrared.
}
\label{nHtoN.ps}
\end{figure}

The standard use of the word ``dispersion'' in the realm of refractive indices
is the prismatic dispersion due to $n'(k)\neq 0$. Linear dispersion
may then describe a linear relation between the refractive index and
wavelength (called \emph{first order dispersion} in \citep{MeisnerSPIE4838,MeisnerSPIE5491})
or between refractive index and wave-number. For the
theory described here, we will obviously use the latter, also keeping
in mind that a Sellmeier functional description near resonances
of Lorentzian shape would support this assumption.
The goodness of fit of the actual air refractive index to either
$\partial n/\partial \lambda\approx$ const or
$\partial n/\partial k\approx$ const is coincidental: whereas 
Fig.\ \ref{nHK.ps} indicates that $\partial n/\partial k$ is approximately
constant in the K band, the equivalent plots at longer infrared
regions are more complicated, see Fig.\ \ref{nHtoN.ps}.

For the wave packet analysis, the focus shifts from $n(k)$ to $n_g(k)$
and its derivatives. Eq.\ (\ref{eq.ngprimeofn}) proves
that a linear dispersion in the sense of $n''(k)=0$ induces
a linear dispersion $n_g''(k)=0$ at the same $k$. However, Fig.\ \ref{n.ps}
demonstrates that the nonlinearity of $n_g$ has a tendency
of being stronger than the nonlinearity of $n$, inherited
from the admixture of higher derivatives of $n$ in the last terms of
Eqs.\ (\ref{eq.ngofn})--(\ref{eq.ngdprime}).
The curves of constant group delay $n_g'(k)=0$ are
of the form $n(k)=f/k+n_\infty$ with constants $f$ and $n_\infty$,
which means they need linearity between $n$ and $\lambda$.

\subsection{Effective Wave-number}
\subsubsection{Choice}\label{sec.kbar}
The fringe packet lives at the DL position $D_i$
where the amplitude $(C^2+S^2)^{1/2}$ of (\ref{eq.Aampl}) is 
maximum, which means $C$ and $S$ should have no linear terms
in the vicinity. The dominating non-constant term
$\propto e_2$ in (\ref{eq.Cdef}) has already a quadratic
form. The dominating leading term $\propto e_1$ in (\ref{eq.Sdef})
has a linear component in $D_i$; from there we may summarize
the rationale of group delay tracking:
\begin{itemize}
\item
enforce
\begin{equation}
e_1=0
\label{eq.e1zero}
\end{equation}
by choosing
\begin{equation}
\bar k= \frac{\int ke(k)dk}{\int e(k) dk} =2\pi\bar\sigma
\label{eq.kbardef}
\end{equation}
as the effective center of the band, then
\item
set $D_i$ according to (\ref{eq.Digroup}) such that
the value of $C$ falls off quadratically from $e_0$; the factor in front
of the term $\propto e_3$ of (\ref{eq.Cdef}) and the cubic factor
in one of the terms $\propto e_3$ of (\ref{eq.Sdef}) become zero.
\end{itemize}
As a by-product, the term $\propto S$ becomes small. With (\ref{eq.effphi}),
$\varphi_i$ becomes close to zero, and the remaining
oscillation $\propto \cos \varphi(\bar k)$ show that the
definition (\ref{eq.kbardef}) has indeed picked up the very
point $\bar k$ of the spectrum where the phase $\varphi$ calculated
from $n$ assumes the value of the entire wave packet represented by $n_g$.
The distances between the lobes are characterized by
$\Delta \varphi(\bar k)=2\pi$
which implies $\Delta D_i=1/[\bar\sigma n(\bar \sigma)]$.

In this approximation, the fringe envelope in (\ref{eq.Aampl}) is
$\surd(C^2+S^2)\approx C\approx e_0-\frac{1}{2}(n_gD_i-D_e)^2e_2$,
which recovers a well-known estimate for the package width:
The half width at half maximum is defined as $C$ drops from $e_0$
to $e_0/2$, which happens at $(n_gD_i-D_e)^2=e_0/e_2$. Therefore 
\begin{equation}
D_i^\mathrm{FWHM}=2\sqrt{e_0/e_2}
\end{equation}
is the full width at half maximum.
If the spectrum is flat over its full width $k^\mathrm{FW}$,
$e(k)=I$ for $|k-\bar k|\le k^\mathrm{FW}/2$,
the moments are $e_0=I k^\mathrm{FW}$ and
$e_2=I\int_{-k^\mathrm{FW}/2}^{k^\mathrm{FW}/2}k^2dk=I(k^\mathrm{FW})^3/12$
to give
\begin{equation}
e_2/e_0 = (k^\mathrm{FW})^2/12, 
\label{eq.e2e0}
\end{equation}
\begin{equation}
D_i^\mathrm{FWHM}=2\sqrt{12}\frac{1}{k^\mathrm{FW}}=\frac{2\sqrt{3}}{\pi}\frac{1}{\sigma^\mathrm{FW}}
\approx \frac{1.1}{\sigma^\mathrm{FW}}.
\label{eq.fwhm}
\end{equation}
This results in a package length of $D_i^\mathrm{FWHM}=14$ $\mu$m
if applied to a full K band width of
$\sigma^\mathrm{FW}\approx 800$ cm$^{-1}=8\times 10^4$ m$^{-1}$,
for example.
Eq.\ (\ref{eq.fwhm}) is the uncertainty relation
for the Fourier-conjugated variables $\sigma$ and $D_i$
in the specific case of a boxcar spectrum.

\subsubsection{Definition}
Note that the variable
\begin{equation}
\bar\lambda
= \frac{\int e(\lambda) \lambda\, d\lambda }{\int e(\lambda)\, d\lambda}
\label{eq.lambarDef}
\end{equation}
is \emph{not} used and that $\bar\lambda \neq 2\pi/\bar k$
because $\lambda=2\pi/k$ means $d\lambda/\lambda =-dk/k$ but
$\lambda d\lambda\neq -k dk$ \citep{MilmanJOSA22}.
The relation is
\begin{eqnarray}
\bar \lambda&=&2\pi\frac{\int \frac{e(k)}{k^3}dk}
{\int \frac{e(k)}{k^2}dk}
=\frac{2\pi}{\bar k}\,
\frac{1-\frac{3}{\bar k}\frac{e_1}{e_0}+\frac{6}{\bar k^2}\frac{e_2}{e_0}+\cdots}
{1-\frac{2}{\bar k}\frac{e_1}{e_0}+\frac{3}{\bar k^2}\frac{e_2}{e_0}+\cdots} 
\nonumber \\
&=& \frac{1}{\bar \sigma}
\left(1-\frac{1}{\bar k}\frac{e_1}{e_0}+\frac{3}{\bar k^2}\frac{e_2}{e_0}+\cdots\right),
\end{eqnarray}
which means with (\ref{eq.e1zero}) that $\bar\lambda > 1/\bar \sigma$
since $e_2$ and $e_0$ are moments of even order and positive,
and that neglect of the term $\propto e_2$ would
underestimate $\bar\sigma$.

With the estimate (\ref{eq.e2e0}) for a flat spectrum,
the relative correction is of the
order of $3e_2/(\bar k^2e_0)\approx (k^\mathrm{FW}/\bar k)^2/4$,
or $1/400$ for a relative band width of $1/10$.
Confirmed numerically, erroneous use of 
$1/\bar\lambda$ for $\bar\sigma$ yields values
that are typically $11$ cm$^{-1}$ off the correct $\bar\sigma$
of the K band center pixel in our wide-band example.

\subsection{Nonlinear Group Delay}

The chirps in the envelopes are caused by linear dispersion (non-vanishing $n_g'$)
and skew correlated spectra, i.e., non-vanishing $e_3$.

The moment expansion has so far recovered the familiar equations
(\ref{eq.Digroup}) and (\ref{eq.kbardef}) on the basis of neglecting
the four terms in (\ref{eq.Cdef})--(\ref{eq.Sdef}) with factors $n_g'$
and/or $e_3$.
The largest term of these is the one $S\propto e_2$. Besides
increasing the amplitude of $A$ \citep{LevequeASS239,LawsonAO35,TangoAO29},
it only has a  tiny effect
on the location of the maximum, adding
$\approx D(\frac{n_g'}{2n_g})^2\frac{e_2}{e_0}$
to $D_i$.
All three neglected terms $S\propto e_2$ or $S\propto e_3$
actually have negligible
effect on the relocation of the envelope maximum, but
the term $C\propto e_3$ \emph{does} have an effect.
As a first new result of the moment expansion we find
a corrected envelope location via
\begin{eqnarray}
\frac{\partial}{\partial D_i}C&=&0, \nonumber \\
\therefore
\frac{\partial}{\partial D_i}
\left\{
\left[n_gD_i-D_e\right]^2e_2+\left[n_gD_i-D_e\right]n_g'D_ie_3
\right\}&=&0. \nonumber
\end{eqnarray}

The solution $D_i$ to this equation slightly modifies (\ref{eq.Digroup}),
\begin{equation}
D_i=\frac{D_e}{n_g}\frac{1+\frac{1}{2}\frac{n_g'}{n_g}\frac{e_3}{e_2}}
{1+\frac{n_g'}{n_g}\frac{e_3}{e_2}}
\approx \frac{D_e}{n_e},
\label{eq.Destim2}
\end{equation}
where an ``extended'' refractive index $n_e$ emerges as
\begin{equation}
n_e(k)\equiv n_g(k)+\frac{e_3}{2e_2}n'_g(k).
\label{eq.next}
\end{equation}
Examples of reaching beyond the group delay approximation in this fashion are
provided in Tab.\ \ref{tab.Destim} related to
Fig.\ \ref{fringe100.ps}\@. The relative correction
read from the last column is of the order of $2\times 10^{-9}$,
generating a correction to the group delay estimate of 200 nm
for a 100 m delay, for example.
The cases of Tab.\ \ref{tab.Destim2}
show that the corrections are reduced to a quarter of these values,
if the band pass is reduced to half of this width.

\begin{table}[hbt]
\begin{tabular}{ccc|cccc}
Star 
   & \multicolumn{2}{c}{atmosphere} 
   & $\bar\sigma$ &
   $e_3/e_2$
   & $n_g'(\bar\sigma)$
   & $\frac{1}{2}\frac{e_3}{e_2}n_g'(\bar\sigma)$ \\
(K) & sec $z$ & (mm) & (cm$^{-1}$) & ($10^3$/m) & ($10^{-13}$ m) & ($10^{-9}$)\\
\hline
8000 & 1 & 1.6 & 4498.4 & $-12.8$ & $3.2$ & $-2.05$ \\
9000 & 1 & 1.6 & 4499.0 & $-13.2$ & $3.2$ & $-2.11$ \\
9000 & 1 & 3.0 & 4497.8 & $-11.9$ & $3.2$ & $-1.90$ \\
15000 & 1 & 3.0 & 4499.6 & $-13.0$ & $3.2$ & $-2.08$ \\
\hline
\end{tabular}
\caption{Change of some variables of the group delay correction
(\ref{eq.next}) as a function of the black body temperature in the star model,
air mass $\sec z$ and precipitable water vapor (PWV) of the atmospheric model, integrated
over a band pass from 2.05 to 2.45 $\mu$m.
\label{tab.Destim}
}
\end{table}
\begin{table}[hbt]
\begin{tabular}{ccc|cccc}
Star 
   & \multicolumn{2}{c}{atmosphere} 
   & $\bar\sigma$ &
   $e_3/e_2$
   & $n_g'(\bar\sigma)$
   & $\frac{1}{2}\frac{e_3}{e_2}n_g'(\bar\sigma)$ \\
(K) & sec $z$ & (mm) & (cm$^{-1}$) & ($10^3$/m) & ($10^{-13}$ m) & ($10^{-10}$)\\
\hline
8000 & 1 & 1.6 & 4459.7 & $-3.77$ & $3.0$ & $-5.66$ \\
9000 & 1 & 1.6 & 4459.8 & $-3.87$ & $3.0$ & $-5.81$ \\
9000 & 1 & 3.0 & 4459.1 & $-3.41$ & $3.0$ & $-5.12$ \\
15000 & 1 & 3.0 & 4459.6 & $-3.70$ & $3.0$ & $-5.55$ \\
\hline
\end{tabular}
\caption{Change of some variables of the group delay correction
(\ref{eq.next}) as in Table \ref{tab.Destim}
over a band pass from 2.15 to 2.35 $\mu$m
(from 4255 to 4651 cm$^{-1}$, $\sigma^\mathrm{FW}=395$ cm$^{-1}$).
$\bar\sigma$ is close to $\sigma_\mathrm{ref}$ of Table \ref{tab.fitn_1}
such that $n_g'$ is close to (\ref{eq.ngprimeref}).
\label{tab.Destim2}
}
\end{table}

Overall, this is a more systematic and consistent approach than to correct the
representative point $\bar k$ in the band for star spectra within the group
delay approximation \citep{DaigneAApS138}.
All effects which change the spectrum---categories listed in Section \ref{sec.emkmodl}---
are treated on equal footing.

The group delay and the correction introduced above are only defined
as some form of interpolation between the fringe peaks; they try to locate
the fringe package envelope to higher precision than the scale defined
by the fringe spacing \citep{MeisnerIAC36}.

\section{ABCD Phase}\label{sec.abcdphase}
\subsection{Achromatic Phase Rotation}\label{sec.phaseErr}
Four
values called A, B, C and D originate from adding phases of $0,\pi/2, \pi$ and $3\pi/2$ to the argument
of the cosine in Eq.\ (\ref{eq.Idfring}),
\begin{equation}
\left.
\begin{array}{c}
A \\
B \\
C \\
D \\
\end{array}
\right\}
=\int e(k)\{1 + \cos[ n(k)kD_t-kD_e +\left\{
\begin{array}{c}
0 \\
\pi/2 \\
\pi \\
3\pi/2
\end{array}
\right\}]\} dk .
\label{eq.F2Destim}
\end{equation}
This writing adds an achromatic
phase to the trigonometric function,
as if an ideal reversion prism (\citep{MorenoAO43,PegisAO2}) had been inserted in one of the
telescope beams in front of the beam combiner to push the relative phase
between the $s$- and $p$-polarized components of its fields by $\pi/2$.

We are not discussing
stepping methods
\citep{BasdenMNRAS357,PadillaSPIE3350,JaffeSPIE5491} which add a ladder
of achromatic paths $sd_i$ ($s=0,1,2,\ldots$) to $D_i$.
The stepping
adds a chirped phase $\propto snkd_i$ to the cosine of (\ref{eq.Idfring}).
In sufficiently narrow bands or when $n(k)$ has a hyperbolic shape,
both types of scanning may essentially be equivalent---putting aside the
question of sensitivity to the polarization of the correlated $e(k)$.

\subsection{Moment Expansion}
\subsubsection{Standard Formula}
We pursue the same moment expansion as in Section \ref{sec.emomen}, up
to the order of $e_2$,
\begin{eqnarray}
A-C && \approx
2e_0\cos \varphi(\bar k) \nonumber \\
&& -2\left\{ n_g(\bar k) D_i-D_e\right\} e_1 \sin \varphi(\bar k) \nonumber \\
&& - \left\{ n_g(\bar k) D_i-D_e\right\}^2
e_2 \cos \varphi(\bar k)
 \nonumber \\
&& - n_g'(\bar k) D_i
e_2 \sin \varphi(\bar k)
 +\cdots , \label{eq.AminC}\\
D-B && \approx
2e_0 \sin \varphi(\bar k) \nonumber \\
&& +2\left\{ n_g(\bar k) D_i-D_e\right\}e_1\cos \varphi(\bar k) \nonumber \\
&& - \left\{ n_g(\bar k) D_i-D_e\right\}^2
e_2 \sin \varphi(\bar k)
 \nonumber \\
&& + n_g'(\bar k) D_i
e_2\cos \varphi(\bar k)
 +\cdots  \label{eq.DminB}.
\end{eqnarray}
When the choice (\ref{eq.kbardef}) is made again, the leading terms $\propto e_0$
dominate $A-C$ and $D-B$.
The result of neglecting all other terms
is known as the ABCD algorithm \citep{ShaoJOSA67,WyantAO14,CaiAO45,ZhongJopt8}:
\begin{equation}
\tan \varphi^\mathrm{ABCD}(\bar k) = \frac{D-B}{A-C}.
\label{eq.abcd}
\end{equation}
The difference to dealing only with the signal $A$ of (\ref{eq.Aampl}) is 
that the information provided in the signals $B$ to $D$ allows to decouple
the data into fringe amplitude and phase.
In the following, complementarily to the contents of section \ref{sec.gDelay}, we
look at the reduction of this \emph{phase tracking approach} to the
value of $D_e$, the ultimate variable of astronomical interest.

In this sense, the construction of the mean vacuum wave vector $\bar k$
of (\ref{eq.kbardef}) selects
that ray, spot or piece in the broad band spectrum which experiences the
geometric path delay $D_e$ above the atmosphere and then the optical
path delay $n(\bar k)D_i$ on the ground such that its residual phase
equals the unwrapped ABCD phase $\varphi$. (Small corrections to this
statement follow in Sect.\ \ref{sec.abcdErr}.) If one describes long baseline
interferometry as the  art of equilibrating the optical path difference $nD_i$
on the ground to the geometric path difference $D_e$ above the atmosphere,
this is only correct for one special $k$ inside the
broad-band spectrum.

\subsubsection{Delay Line Mirror Vibrations}
The implicit integration over varying delays caused
by mirror surface vibrations on shorter timescales than the detector integration
adds explicit
time-dependence to the positions $D_i$ in the arguments of the
trigonometric functions (\ref{eq.Idfring}). On short timescales,
these are to lowest order linear, $D_i\rightarrow D_i+2vt$, where
$v$ is the
velocity of the mirror surface. First-order
expansion of the trigonometric function $\propto t$ yields no change
because the integral over the exposure time, $-t_e/2\le t\le t_e/2$,
vanishes. Including the second order, the trigonometric functions are multiplied
by the factor $1-(2nkvt)^2/2$, which looks like the lowest order of
the familiar exponential Strehl factor of phase variance $2nkvt$.

The integral over the time interval $t_e$ result in a multiplication of $e(k)$
by $1-v^2t_e^3k^2/6$.
(Vibrations which are sensed by a metrology generate other
effects \citep{ComolliRevSci76}.)
Expanding this term around some $\bar k$ generates a
common factor for $e_0$ to all four ABCD values of a band; it reduces the
measured visibility, but cancels when building the estimator
$\varphi^\mathrm{ABCD}$ with (\ref{eq.abcd}).
The change of $\bar k$ and contribution to the moment $e_2$, however, is
one of the effects that
are covered implicitly by the present analysis.
(Harmonic mirror vibrations could be treated as Strehl factors \citep{AltaracMNRAS322}
emulated by Bessel Functions, which we do not detail further.)

\subsection{Nonlinear ABCD Phase}\label{sec.abcdErr}
\subsubsection{Wideband phase correction}
If the first moment $e_1$ vanishes (see Section \ref{sec.kbar}),
the system of equations (\ref{eq.AminC})--(\ref{eq.DminB})
has the matrix-vector form
\begin{eqnarray}
&&\left(
\begin{array}{c}
A-C \\
D-B \\
\end{array}
\right)
\equiv
\left(
\begin{array}{c}
X \\
Y \\
\end{array}
\right)
= 
\label{eq.XYmatr}
\\
&&\left(
\begin{array}{cc}
\cos\varphi(\bar k) & -\sin\varphi(\bar k) \\
\sin\varphi(\bar k) & \cos\varphi(\bar k) \\
\end{array}
\right)\cdot
\left(
\begin{array}{c}
2e_0-[n_g(\bar k)D_i-D_e]^2e_2 \\
n_g'(\bar k)D_ie_2 \\
\end{array}
\right). \nonumber
\end{eqnarray}
[In the interpretation of $A-C$ as the $p$-component and the $D-C$
as the $s$-component of the beams, the correction by a (small)
$n_g'D_ie_2$ appears as a fake polarization.]
We multiply this equation with the inverse of the rotation matrix,
introduce the dimensionless parameter
\begin{equation}
\zeta\equiv \frac{n_g'D_ie_2}{2e_0-[n_gD_i-D_e]^2e_2},
\label{eq.r}
\end{equation}
and find, with $\tan\varphi^\mathrm{ABCD}=Y/X$,
\begin{equation}
\tan\varphi(\bar k)=\frac{Y-X\zeta}{Y\zeta+X}
=
\frac{\tan\varphi^\mathrm{ABCD}-\zeta}{1+\zeta\tan\varphi^\mathrm{ABCD}}.
\label{eq.tanphizeta}
\end{equation}
Taylor expansion of the arctangent of this equation for small $\zeta$ yields
the corrected phase
\begin{eqnarray}
\varphi(\bar k)&=&\varphi^\mathrm{ABCD}-\arctan(\zeta) \nonumber \\
&\approx& \varphi^\mathrm{ABCD}-\zeta+\frac{1}{3}\zeta^3+O(\zeta^5).
\label{eq.phicorr}
\end{eqnarray}
The sign of the correction demonstrates that the observed $\varphi^\mathrm{ABCD}$
is larger than the phase $\varphi(\bar k)$ assigned to the monochromatic reduction,
if $n_g'>0$. This is expected because $e_2$ describes adding spectral elements
symmetrically to both sides of $\bar k$, and a positive derivative $n'$
adds a term $\propto k^2$ to the phase (\ref{eq.phi}), which
un-balances the otherwise symmetric weighting by $e_2$ in
the statistics of the phase spectrum (\ref{eq.phi}).

The prototypical values for the example of spectra integrated over
from 4081 to 4878 cm$^{-1}$ (2.05 to 2.45 $\mu$m)
of black body temperature from 6000 to 12\,000 K are $e_2/e_0\approx 4.9\times 10^4$
cm$^{-2}$
[larger than the flat spectrum estimate of (\ref{eq.e2e0}) at
$\sigma^\mathrm{FW}\approx 800$ cm$^{-1}$],
$n_g'\approx 3.2\times 10^{-11}$ cm from Table \ref{tab.Destim},
and for $D_i=100$ m
\begin{equation}
\zeta\approx \frac{1}{2}\frac{e_2}{e_0}n_g'D_i\approx 0.008\, \mathrm{rad}.
\label{eq.rappro}
\end{equation}

This phase correction of 0.008 rad equals a change in $D_e$
of $0.008/(2\pi\bar\sigma)\approx 2.8$ nm at $\bar\sigma\approx 4500$ cm$^{-1}$.
We observe that this correction from $\varphi^\mathrm{ABCD}$
to $\varphi(k)$ is only 1.5 \% of the one calculated for the envelope shift
after (\ref{eq.next}).
The wide band correction shifts the fringe envelope along the
$D_i$ abscissa; the individual fringe lobes stay almost pinned.
However, this result may be deceptive since the estimate (\ref{eq.rappro})
of (\ref{eq.r})
assumes that the term $\frac{1}{2}\frac{[n_gD_i-D_e]^2e_2}{e_0}$
in the denominator of (\ref{eq.r}) can be neglected.
Since $e_2/(2e_0)\approx 2/(D_i^\mathrm{FWHM})^2$ as estimated
in Section \ref{sec.kbar},
this assumption is only valid if the tracking settles $D_i$ close
to the envelope maximum, at $n_gD_i-D_e\approx 0$. Tracking a distance
of $\frac{1}{2} D_i^\mathrm{FWHM}$ away from the maximum,
$\frac{1}{2}\frac{[n_gD_i-D_e]^2e_2}{e_0}$ is approximately $\frac{1}{2}$,
and $\zeta$ in (\ref{eq.rappro}) grows by a factor
of 2, for example.
So precise spectrum correction in the phase tracking needs some
sort of group tracking assistance to produce the estimator of $n_gD_i-D_e$
to feed the estimation of $\zeta$; if the expansion
(\ref{eq.phicorr}) breaks down because the correction $\zeta$ becomes too
large,
one must turn to a self-consistent solution of (\ref{eq.r}) and
(\ref{eq.tanphizeta}) where $\zeta$ and $\varphi(\bar k)$
are both functions of the unknown $D_e$.

\subsubsection{Imprecise Spectra}
A further source of systematic error is that $\bar k$ (the effective
star color) is not known exactly but only with an error $\Delta\bar k$,
which creates a residual $e_1\approx e_0\Delta\bar k$.
The terms $\propto e_1$ in $D-B$ and $A-C$ do not vanish
exactly, and (\ref{eq.AminC})--(\ref{eq.DminB}) become
\begin{eqnarray}
X&=&
2e_0\cos\varphi(\bar k)-2e_0\Delta\bar k[n_gD_i-D_e]\sin\varphi(\bar k)+\cdots, \nonumber \\
Y&=&2e_0\sin\varphi(\bar k)+2e_0\Delta\bar k[n_gD_i-D_e]\cos\varphi(\bar k)+\cdots . \nonumber
\end{eqnarray}
The format still fits into (\ref{eq.XYmatr}) if we replace
$n_g'D_ie_2\leadsto
n_g'D_ie_2+2e_0\Delta\bar k(n_gD_i-D_e)$.
The induced correction in the phase (\ref{eq.rappro}) is
\begin{equation}
\Delta \zeta\approx \Delta \bar k(n_gD_i-D_e).
\end{equation}
Assuming again phase tracking allowing some
ambiguity in the ``best'' lobe (``fringe jumps''), $n_gD_i-D_e$
will be of the order of $2\lambda$, approximately $4\times 10^{-4}$ cm
in the K band. The required $\Delta k<\Delta \zeta/(n_gD_i-D_e)$
follows from an allowable error $\Delta \zeta$ in the phase: a 10 \% error
in the phase, $\Delta \zeta=0.63$ rad, would admit an error
of $\Delta\bar k\approx\frac{0.63}{2\lambda}\approx \frac{0.63}{4\times 10^{-4}\mathrm{cm}}$,
which is $\Delta\bar\sigma \approx 250$ cm$^{-1}$.
This is a quarter of the full band width and therefore easily met.

So on behalf of this particular wide-band correction to the
ABCD phase, knowledge of the mean momentum number $\bar k$ is not a concern;
the role of $\bar k$ in selecting $n(\bar k)$ is a different theme
and turns out to impose stronger error bounds on $\bar k$,
as we shall see in Section \ref{sec.kjitter}.

\section{Astrometric Application}\label{sec.kjitter}
\subsection{Data Reduction}
If the fringe tracker locates the fringe package maximum,
the data reduction converts the delay line mirror position $D_i$
at that point into $D_e=n_e(\bar k)D_i$ of (\ref{eq.Destim2}).
If the tracker senses the ABCD phase, the data reduction applies
the correction (\ref{eq.phicorr}) and reduces (\ref{eq.Idfring}) to
\begin{equation}
D_e=n(\bar k)D_i
-\frac{\varphi(\bar k)}{\bar k}
\label{eq.Deofkbar}
\end{equation}
of (\ref{eq.phi}).
The task of disentangling the two terms in the cosine
of (\ref{eq.Idfring}) is a key difference between
an astrometric data analysis and a fringe tracker.

In both cases, the delay line positions $D_i$ are probably obtained
from an auxiliary metrology system (which we did not address),
and in both cases of the product $n_{(e)}(\bar k)D_i$ enters
the metrology reading such that it reduces the errors if the
metrology is already operating close to $\bar k$.
A mixed/hybrid operation of both methods is useful.
The common well-known problem of both approaches is the fluctuation
of $\bar k$ as a function of the spectral shape of $e(k)$, which
enters the data reduction directly as $\bar k$ or indirectly
as $n_e(\bar k)$, which we discuss in the final subsections.

\subsection{Unknown Star Temperature}

\subsubsection{Sensitivity}
This well-known hybrid sensitivity
of the central $\bar k$ to the spectral width and star color is estimated
in Tab.\ \ref{tab.chWids}.
Na\"{\i}ve scaling of the sensitivity proportional
to the inverse squared width of the spectral channel turns out
to be quite inaccurate.

\begin{table}[hbt]
\begin{tabular}{c|cc}
$\lambda$ range ($\mu$m) & $\bar{\sigma}$ (8000 K) & $\Delta \bar{\sigma}$ ($\leadsto$ 9000 K) \\
\hline
2.05 -- 2.45 & 4498.0 & 0.656 \\
\hline
2.05 -- 2.26 & 4657.0 & 0.215 \\
2.26 -- 2.45 & 4256.8 & 0.121 \\
\hline
2.05 -- 2.19 & 4724.7 & 0.102\\
2.19 -- 2.32 & 4440.2 & 0.068\\
2.32 -- 2.45 &  4197.6 & 0.054\\
\hline
2.05 -- 2.15 & 4766.0 & 0.054 \\
2.15 -- 2.25 & 4549.0 & 0.045\\
2.25 -- 2.35 & 4350.9 & 0.037\\
2.35 -- 2.45 & 4169.4 & 0.031\\
\end{tabular}
\caption{
The sensitivity of the pivot wave number $\bar{\sigma}$ to a change of
the temperature of a pure black-body spectrum for
different spectral resolution (subdivisions) of the K band.
In accordance with Fig.\ \ref{blackb_tab.ps}, the mean wave number shifts
to higher values for the increase from 8000 to 9000 K.
}
\label{tab.chWids}
\end{table}

Since Fig.\ \ref{spektrk.ps} will show
that the curves become flatter at higher temperatures, the example
at 8000 K is already some worst case.

\subsubsection{Phase Tracking}
The astrometric use of these results would concentrate on the estimate
of $D_e$ and derive it from the measured phase via (\ref{eq.Deofkbar}).
The ``direct scaling'' error in $\varphi/\bar k$ from a poorly known star
temperature $T$ is the first order differential
\begin{equation}
\Delta D_e\approx
\frac{\varphi}{\bar k^2}\Delta \bar k
=
\frac{\varphi/(2\pi)}{\bar \sigma^2}\,\frac{\partial \bar \sigma}{\partial T}\Delta T.
\label{eq.DeDeltaT}
\end{equation}
The term $\frac{\partial \bar \sigma}{\partial T}\Delta T$
would be of the order of 0.7 cm$^{-1}$ for an error $\Delta T\approx 1000$ K,
see the first line in Table \ref{tab.chWids}\@.
At 100 m of delay, the term $\varphi/(2\pi)$ would be $\approx -29$
cycles, see the lower graph in Fig.\ \ref{phaseH.ps}\@.
Using in addition $\bar\sigma\approx 4500$ cm$^{-1}$, the ``direct scaling'' error
(\ref{eq.DeDeltaT}) is $\Delta D_e\approx -10$ nm.

\subsubsection{Group Delay Tracking}
If the data reduction uses (\ref{eq.Destim2}) with (\ref{eq.next}),
the precise value of $\bar k$ determines where on
the humid air dispersion curve $n$ has to be read.
Since $\bar k$ represents a spectrum, the number of spectral channels
to represent the band takes influence on this knowledge
of the differential index of refraction, as a higher number of spectral
channels means less sensitivity to the star colors---this is essentially
the same result as from comparison of Figs.\ 12 and 13 in \citep{MilmanJOSA22}.
Each change of $\bar{\sigma}$
by 1 cm$^{-1}$ induces a change of $n$
by $1.1\times 10^{-10}$, see $c_{1\mathrm{ref}}$ of Table \ref{tab.fitn_1}\@.
With the same error in $\bar\sigma\approx 0.7$ cm$^{-1}$ derived from
$\Delta T\approx 1000$ K as in the previous subsection, the
error in $n$ reaches $\approx 8\times 10^{-11}$, which is $+8$ nm of
``indirect'' error per 100 m of delay.

This sensitivity to a change in $\bar\sigma$ induced by a change in the star
color is practically the same as in the previous section with the ``Phase
Tracking'' reduction. [This is no coincidence.
The condition (\ref{eq.phiDeriv})
actually sets up $\varphi$ to mediate this scaling:
Consider the
derivative $\frac{\partial(\varphi/\sigma)}{\partial \sigma}
=\frac{\partial\varphi/\partial \sigma}{\sigma}-\frac{\varphi}{\sigma^2}$
which becomes $-\frac{\varphi}{\sigma^2}$ via (\ref{eq.phiDeriv}).
Using (\ref{eq.phi}), the derivative can also be written
$2\pi\frac{\partial(nD_i-D_e)}{\partial \sigma}
\approx 2\pi D_i\frac{\partial n}{\partial \sigma}$.
This allows the substitution
$\frac{\varphi/(2\pi)}{\bar\sigma^2}
 \leadsto -D_i\frac{\partial n}{\partial\bar\sigma}$
in (\ref{eq.DeDeltaT}) to prove equivalence of both error propagations.]

\subsection{Varying Air Mass and PWV}\label{sec.pwv}
The influence of variations in the precipitable water vapor (PWV) and air mass
on the mean wave number in the spectral channels is investigated
by considering Fig.\ \ref{trans.ps} for water vapor columns of 1, 1.6 and 3 mm,
and for air masses of 1 and 1.5\@. (See \citep{CormierJCP122} for a discussion
of water vapor absorption in the infrared.)

\begin{figure}[hbt]
\includegraphics[width=8.5cm]{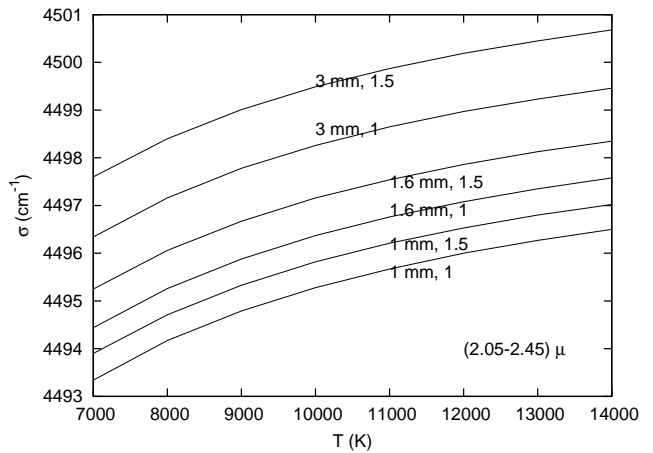}
\caption{
Black body spectra from 7\,000 to 14\,000 K multiplied with the air transmission
spectrum for 6 pairs of PWV and air mass and multiplied
with the detector quantum efficiency define these effective wave numbers in
the 2.05--2.45 $\mu$m band according to (\ref{eq.kbardef}).
}
\label{spektrk.ps}
\end{figure}
\begin{figure}[hbt]
\includegraphics[width=8.5cm]{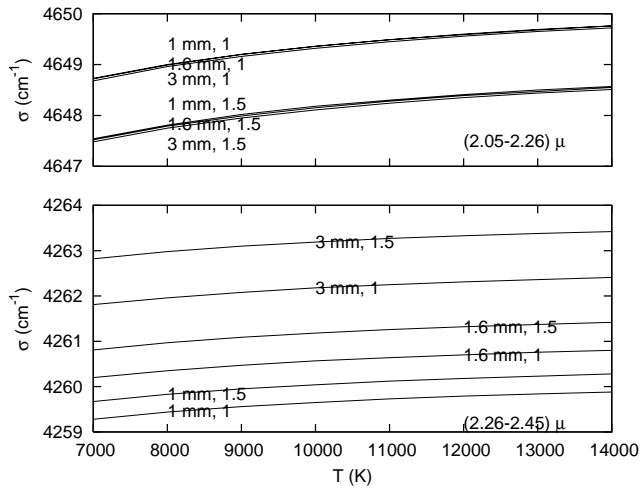}
\caption{
The wide band of Fig.\ \ref{spektrk.ps} is split into
two bands from 2.05 to 2.26 $\mu$m and from $2.26$ to $2.45$ $\mu$m,
leading to these new central wave numbers $\bar \sigma$ as a function
of star temperature, PWV and air mass.
}
\label{spektrk2.ps}
\end{figure}

\begin{figure}[hbt]
\includegraphics[width=8.5cm]{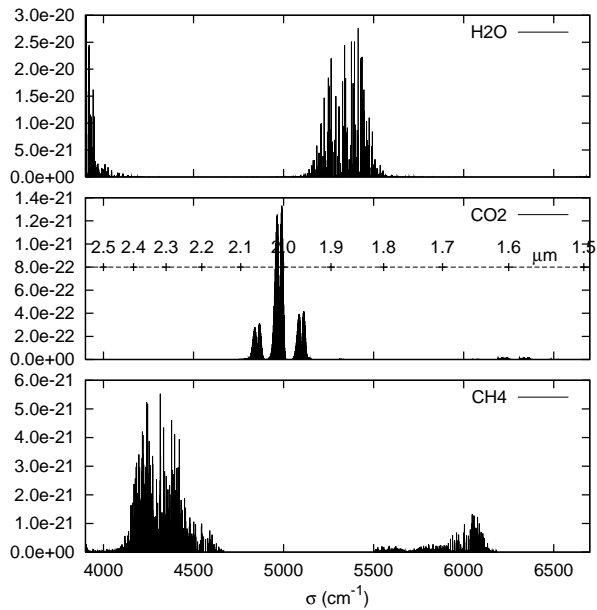}
\caption{
Line strengths of important atmospheric gases (water, carbon dioxide, methane)
in units of cm$^{-1}$/(molecule cm$^{-1}$),
contributing to Fig.\ \ref{spektr.ps} \citep{RothmanJQSRT82}.
}
\label{HitL1.ps}
\end{figure}
Black body spectra
from 1.52 $\mu$m to 2.78 $\mu$m
have been sliced into 20\,000 virtual sub-channels,
multiplied by the air transmission factors and with the detector response
of Fig.\ \ref{256QE.ps} to yield representations $e(k)$ as in
Fig.\ \ref{spektr.ps}\@.
Summation of (\ref{eq.kbardef}) over sub-channels
creates Figs.\ \ref{spektrk.ps} and \ref{spektrk2.ps}\@.
The flatter lines in Fig.\ \ref{spektrk2.ps} show that the variation of the
central $\bar k$ as a function of the black body temperature
is reduced efficiently if the band width is reduced---consistent
with the calculations on pure black body spectra of Table \ref{tab.chWids}\@.

The air mass controls the strength of the CO$_2$ absorption bands
near 4850 cm$^{-1}$, and the PWV controls the  termination of the
atmospheric window near 4200 cm$^{-1}$ (Fig.\ \ref{HitL1.ps}).
Their asymmetric position
in opposite corners in the two bands explains the opposite signs of
the gradients of $\bar\sigma$ with respect to these two parameters
in the two plots of Fig.\ \ref{spektrk2.ps}. For this reason, splitting
the original wide band does not desensitize $\bar\sigma$ from
the sky transmission parameters as easily as from the
broad star spectra.

\section{Summary}
The effect of air dispersion on visibilities measured by
two-beam astronomical interferometers
is accompanied by effects of astrometric signature:
Accumulation of dispersed phases over correlated
wide-band
spectra relocates the
maximum of the wave packet envelope as if the group refractive index
$n_g$ had been replaced by an extended refractive index $n_e$
which equals the group refractive index
$n_g$ augmented by a hybrid product between some spectral skewness
and the dispersion $\partial n_g/\partial k$ of the group refractive index.
The complementary influence on the phase calculated with the ABCD formula
is typically much smaller.

Not only the star spectrum but also the (variable) atmospheric conditions
and other instrumental effects shape the source spectrum and play important
roles in determining pivot wave-numbers of the spectral bands.

\begin{acknowledgments}
This work is supported by the NWO VICI grant of 15-6-2003
``Optical Interferometry: A new Method for Studies of Extrasolar Planets'' to A. Quirrenbach.
\end{acknowledgments}

\bibliography{all}

\end{document}